\documentclass{rmaa}

\title{Expansion Parallax for the Compact Planetary
Nebula M2-43}

 \author{Lizette Guzm\'an, Yolanda G\'omez, and Luis F. Rodr\'\i guez 
  \affil{Centro de Radioastronom\'{\i}a y Astrof\'{\i}sica, UNAM, Morelia}
 }

\fulladdresses{
\item Lizette Guzm\'an, Yolanda G\'omez,
Luis F. Rodr\'\i guez: Centro de Radioastronom\'{\i}a
y Astrof\'\i sica, 
UNAM, A. P. 3-72, 
(Xangari), 58089 Morelia, Michoac\'an, M\'exico
(l.guzman, y.gomez, l.rodriguez@astrosmo.unam.mx).
}

\shortauthor{Guzm\'an et al.}
\shorttitle{Expansion Parallax for the Planetary
Nebula M2-43}

\SetVolume{50} \SetFirstPage{001} \SetYear{2004}
\ReceivedDate{\today} 
\AcceptedDate{Year Month Day} 

\resumen{Presentamos observaciones de alta calidad en
radiocontinuo hechas a 3.6 cm con el Very Large Array
en dos \'epocas hacia la nebulosa planetaria M2-43. La comparaci\'on 
de las dos \'epocas, obtenidas con una separaci\'on de 4.07 a\~nos,
muestra claramente la expansi\'on de la nebulosa planetaria con
una velocidad angular de 0.61 $\pm$ 0.09 mas~a\~no$^{-1}$. Suponiendo que 
la velocidad de expansi\'on
en el plano del cielo (determinada de estas mediciones)
y la velocidad de expansi\'on en la l\'\i nea de
visi\'on (determinada a partir de espectroscop\'\i a
\'optica disponible en la literatura) son iguales, encontramos una distancia
de 6.9 $\pm$ 1.5 kpc a la nebulosa planetaria. Esta es la mayor distancia
a una nebulosa planetaria medida hasta ahora con esta t\'ecnica. 

}
 
\abstract{We present high quality
radio continuum observations made with the Very Large Array
at 3.6 cm at two epochs toward the planetary nebula M2-43.
The comparison of the two epochs, obtained with a time
separation of 4.07 years, clearly shows the expansion
of the planetary nebula with an angular rate of 
0.61 $\pm$ 0.09 mas~year$^{-1}$. 
Assuming that the expansion velocity in the plane of the
sky (determined from these measurements) and the expansion velocity
in the line of sight (determined from optical spectroscopy available in
the literature) are equal, we find a distance to the planetary nebula of  
6.9 $\pm$ 1.5 kpc.  This is the largest distance for a planetary nebula
measured up to now with this technique.  
}
      

\begin{document}

\maketitle

\section{Introduction}

The distance to a planetary nebula (PN) is an essential parameter for studying 
the stellar and nebular parameters as well as the evolution of the central star. 
The measurement of the angular expansion parallax of a PN,
from radio interferometric data obtained over a period of a few years,
provides an accurate method to estimate their distances (Masson 1986). 
This technique has
been applied with success in several PNe (Masson 1986;
1989a; 1989b; G\'omez, Rodr\'\i guez \& Moran 1993; Hajian, Terzian \& Bignell 1993; 1995; 
Kawamura \& Masson 1996; Hajian \& Terzian 1996; Christianto \& Seaquist 1998). The angular
expansion technique has also been used for Hubble Space Telescope WFPC2 data in several
PNe (Reed et al. 1999; Palen et al. 2002). 

In this paper we present an angular expansion study of the planetary 
nebula M2-43, made with Very
Large Array (VLA) data 
taken at 3.6 cm with a time baseline of 4.07 years.
Our main purpose was to detect for the first time this expansion
and to use it to obtain an accurate distance estimate to this PN.
The planetary nebula M2-43 (=PN G027.6+04.2) has been detected at radio wavelengths,
showing a compact size with a diameter of 1${\rlap .}^{\prime\prime}$5 for its major axis
 and ellipsoidal 
morphology (Aaquist \& Kwok 1990).
High-resolution radio free-free (2~cm) emission and hydrogen recombination line (H$\alpha$)
images toward M2-43 show a similar structure with two bright peaks in the north-south 
direction (Lee \& Kwok 2005). The extinction map structure derived by Lee \& Kwok (2005) 
roughly follows that of the radio map.  
Expansion velocities for M2-43 measured with different ions
are in the range from 26 to 30 km~s$^{-1}$ (Acker et al. 2002; 
Pe\~na, Medina \& Stasi\'nska 2003). Acker et al. (2002) found
spectral evidence for turbulent velocities in [WC]-type PNe superimposed on a
constant expansion velocity pattern. The modelling made by Acker et al. (2002)
toward M2-43, taking into account the turbulence, gives an expansion velocity of 
20 km~s$^{-1}$ with a turbulent component of 10 km~s$^{-1}$. 
In addition,  M2-43 has a high radial velocity value (with respect to the 
local standard of rest) of +111.6 $\pm$ 5 km~s$^{-1}$ (Schneider et al. 1983).
Previous estimates of the distance to M2-43 have been made using statistical methods and
range from 1.4 kpc (Cahn et al. 1992) to 5 kpc (Acker et al. 2002).
The central star of this PN has been classified as a Wolf-Rayet type [WC8],
showing pure He and C in its atmosphere and its visual magnitude
is m$_V$=15.7 (Acker et al. 1992; Leuenhagen \& Hamann 1998; Pe\~na, Medina \& 
Stasi\'nska 2003). 
An accurate distance value for M2-43 is needed to better know the physical 
parameters, evolutionary
stage and to understand the difference between PNe with [WC] nuclei and normal PNe.

\section{Observations}

The observations were made in 1995 August 24 (epoch 1995.65)
and 1999 September 19 (epoch 1999.72)
with the VLA of the NRAO\footnote{The National Radio 
Astronomy Observatory is operated by Associated Universities 
Inc. under cooperative agreement with the National Science Foundation.}
at 3.6 cm in the A configuration. The time interval between observations
was of 4.07 years.
In both epochs the source 1328+307 was used as an absolute amplitude
calibrator (with an adopted flux density of 5.21 Jy) and the
source 1741-038 was used as the phase calibrator
(with bootstrapped flux densities of 3.85$\pm$0.04 and 4.94$\pm$0.03 Jy for the 
first and second epochs, respectively). 
The data provided an angular resolution
of $\sim 0\rlap.{''}3$ with natural weighting. 
The data were reduced using the standard VLA procedures and then cross-calibrated
in phase and amplitude using the procedure of Masson (1986; 1989).
Since the (u,v) coverage and the integration time were very similar for
both epochs, the synthesized beams differed by less than 5\% and images for
the two epochs were made with a restoring beam of $0\rlap.{''}32 \times 0\rlap.{''}31$
and a position angle of $0^\circ$ for the major axis, the average of the individual
beams for each epoch. In Figure 1 we show the individual images for each
epoch as well as the difference image, made from the 1999.72 uv-data from which the
clean components of the 1995.65 image were subtracted directly in the uv-plane. This 
difference image clearly shows the signature of expansion
with the outer parts of the source appearing as positive and the 
inner parts of the source appearing as negative. A consistent result is obtain from 
subtracting the individual cross-calibrated images in the image-plane.

\begin{figure*}
\centering
\includegraphics[scale=0.80, angle=0]{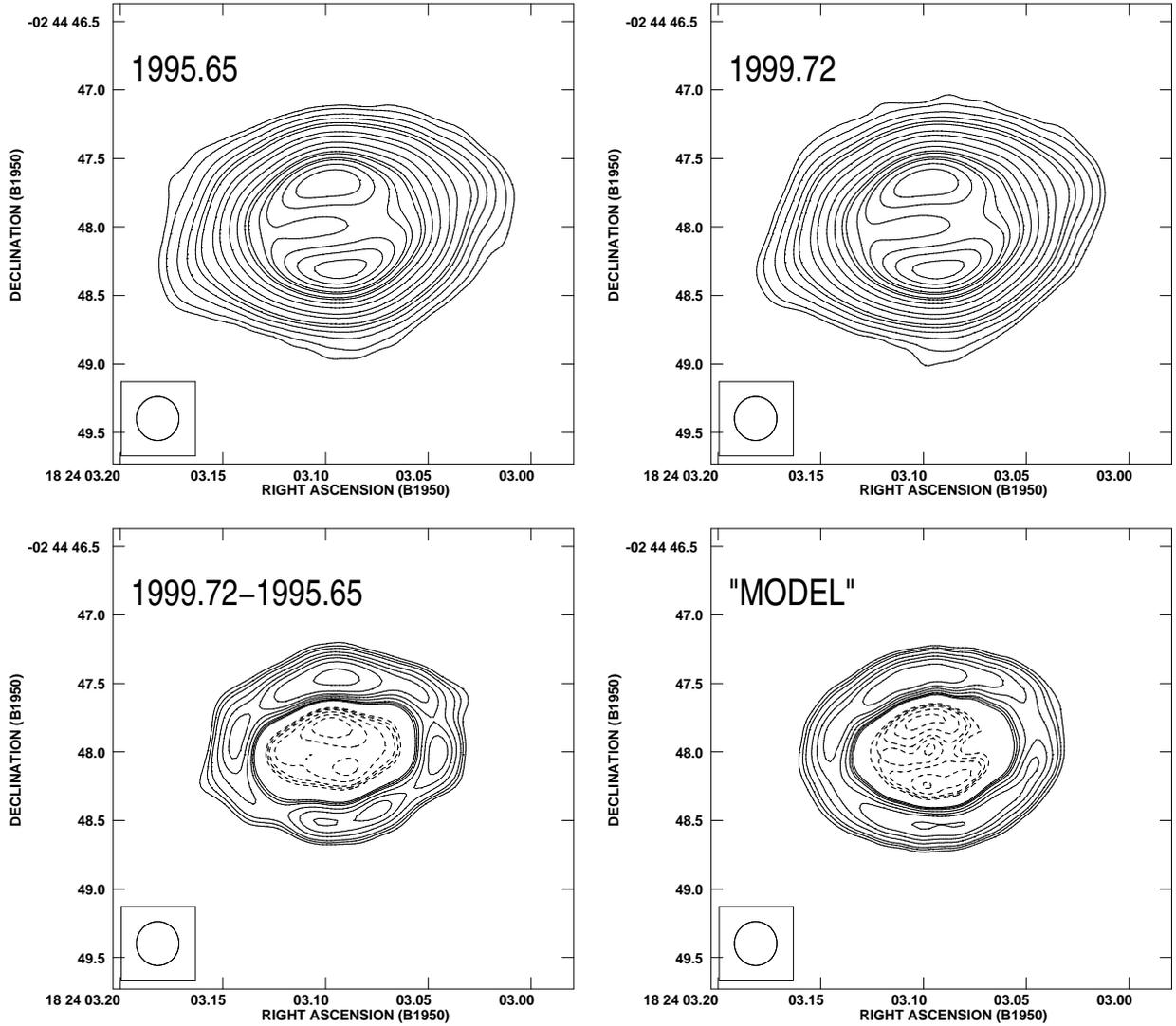}
 \caption{Top: contour images of the 3.6 cm continuum
emission from M2-43 for 1995.65 (left) and 1999.72 (right).
The contours are  -4, 4, 8, 20, 40, 70, 100, 200, 300, 500, 700, 900, 1000,
1100, 1200, 1300, 1500, 1700, and 1900 times 14 $\mu$Jy~beam$^{-1}$, 
the average rms noise of the images. 
Bottom: contours of the 3.6 cm difference image
(left) and of the ``model'' (right) obtained as described in the text.
The contours are  -12, -10, -8, -6, -5, -4, 4, 5, 6, 
8, 10, 12, 15, and 20 times 20 $\mu$Jy~beam$^{-1}$, the rms noise of the 
difference image.
The restoring beam ($0\rlap.{''}32 \times 0\rlap.{''}31$
with a position angle of $0^\circ$) is shown in the bottom left corner
of each image.
}
  \label{fig1}
\end{figure*}

\section{Interpretation and Results}

We modeled the difference map (Fig. 1: bottom-left) following G\'omez, 
Rodr\'\i guez, \& Moran (1993).
A first image was made concatenating the two epoch data sets. A second image was 
made taking the first image and
expanding it in a self-similar way by a factor of
(1 + $\epsilon$), where $\epsilon << 1$. 
The first image was then 
subtracted from the second one for different values of $\epsilon$
to produce a set of ``model'' images.
From a $\chi^2-$square fitting we find that the best agreement with the 
difference image is obtained for
$\epsilon = 0.0075 \pm 0.0008$ (see Fig. 1: bottom-right). 
The good agreement between the difference image and the
model suggests that the assumption of self-similar expansion in M2-43 is 
reasonable. 
The radius of maximum emission, $\theta$, is estimated to be $0\rlap.{''}33 \pm
0\rlap.{''}03$ from the images of the full free-free emission (top part of
Fig. 1). The angular expansion rate of this radius of maximum emission is:

$$ \dot{\theta} = {{\theta~ \epsilon} \over {\Delta t}},$$

\noindent and from our measurements we find 
$\dot{\theta} = 0.61 \pm 0.09~ mas~yr^{-1}$.
The distance to the planetary nebula will then be given by,

$$\bigg[\frac {D} {pc}\bigg] = 211 \bigg[\frac {v_{exp}} {km~s^{-1}}\bigg] \bigg[ \frac {\dot{\theta}} {mas~yr^{-1}} \bigg]^{-1},$$

\noindent where $v_{exp}$ is the expansion velocity of the nebula at the point of
maximum emission. In general the 
[OIII] line widths are used to determine the expansion velocities in PNe. 
Pe\~na et al. (2003)
measured the line width of the [OIII] line at 1/10 the maximum intensity. For M2-43
they obtain a value of 52 km~s$^{-1}$ where not only expansion but also turbulence and high velocity components
are contributing. Acker et al. (2002), found
spectral evidence for turbulent velocities in [WC]-type PN superimposed on a
constant expansion velocity pattern. The modelling made by Acker et al. (2002),
taking into account the turbulence, gives an expansion velocity for M2-43 of 
20 $\pm$ 3 km~s$^{-1}$.

Adopting an expansion velocity of 20 km~s$^{-1}$ for M2-43
in combination with our angular expansion rate we estimate a distance of 
6.9 $\pm$ 1.5 kpc for M2-43. This value is roughly consistent with the distance value
determined by Acker et al. (2002) of 5 kpc. 

Expansion parallax distances using this radio technique have been determined now 
for nine planetary nebulae (Mellema 2004; G\'omez et al. 1993).  
M2-43 is the most distant object in this list, followed by Vy~2-2 for which
a distance of 3.6 $\pm$ 0.4 kpc is estimated (Christianto \& Seaquist 1998).

This value for the distance to M2-43 is consistent with
the fact that it has a large radial LSR velocity of +111.6 $\pm$ 5 km~s$^{-1}$ 
(Schneider et al. 1983). The deviation attributable to velocity dispersion 
with respect to their respective local standard of rest is
around 40 km~s$^{-1}$ for PNe and 20 km~s$^{-1}$ for HII regions 
(Maciel \& Dutra 1992). 
Then, assuming that the LSR radial velocity of the position where
M2-43 is located is +111.6 $\pm$ 40 km s$^{-1}$, we can plot this
velocity and the distance estimated by us (see Figure 2)
to show that both estimates are consistent with a relatively large
distance for M2-43.

\begin{figure}
\centering
\includegraphics[scale=0.40, angle=0]{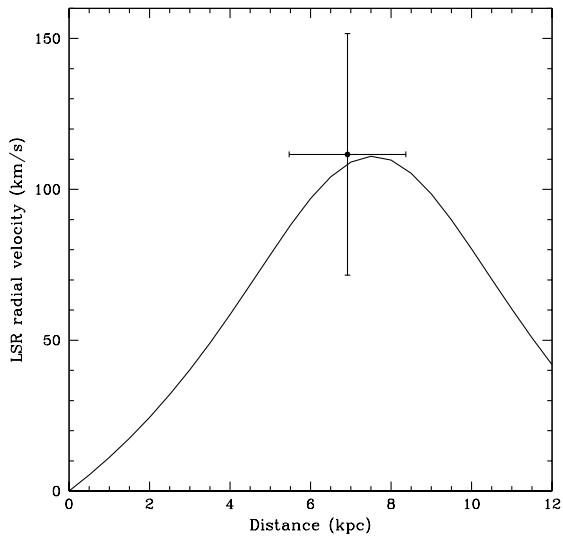}
 \caption{In this figure we show (solid line) the expected LSR radial velocity
as a function of distance to the Sun in the direction
of M2-43 for the galactic rotation model of
Brand \& Blitz (1993). The point marks the estimated distance and the
observed LSR radial velocity for M2-43.
}
  \label{fig2}
\end{figure}

It is clear that the main limitation of this technique
for the distance determination comes from the knowledge of the expansion velocity.
Additional uncertainty in the distance determination is produced by the 
advance of the ionization front and the nonspherical geometry of the nebula. 
Recently, several authors have 
discussed the parallax method for distance determination by modelling the
kinematics of the circumstellar envelopes, proposing correction factors between 
1.3 and 3 for the distance estimates (Mellema 2005; 
Sch\"onberner, Jacob, \& Steffen 2005). However, the application of these correction
factors requires of a very detailed knowledge of the structure of the PN that is not
available at present for M2-43.

The total flux density derived from the 3.6 cm image (second epoch) is 248 $\pm$ 1 mJy and
the mean deconvolved angular diameter at FWHM is 0${\rlap .}^{\prime\prime}$61.
Adopting a distance of 6.9 kpc for M2-43 and assuming that the source is a uniform sphere and
that the radio continuum emission is optically thin at 3.6 cm, we derive an 
emission measure of 4.5 $\times$ 10$^8$ cm$^{-6}$~pc,
an ionized nebular mass of $\sim$ 0.035 M$_\odot$ and an electron density of $\sim$
1.1 $\times$ 10$^5$ cm$^{-3}$. 
An estimate of the kinematic age for M2-43 ($\theta$/$\dot{\theta}$ ) is of 
$\sim$500 years
indicating that this compact PN is very young.

\section{Conclusions}

We presented VLA observations made at 3.6 cm of 
the planetary nebula M2-43 during two epochs separated
by 4.07 years. 
Assuming a self-similar expansion for the
ionized gas we determine an expansion angular rate for M2-43
of 0.61 $\pm$ 0.09 mas~year$^{-1}$ that in combination with an
expansion velocity of 20 km~s$^{-1}$ imply a distance to the
nebula of 6.9 $\pm$ 1.5 kpc. This distance estimate is in agreement
with the high radial LSR velocity reported for this nebula. Adopting a
distance of 6.9 kpc we derive a ionized mass of 0.035 M$_\odot$ and 
a kinematic age of 500 years for M2-43, indicating that it is a very
young and relatively remote planetary nebula.


\acknowledgments
LG, YG and LFR acknowledge the support
of DGAPA, UNAM, and of CONACyT (M\'exico).
This research has made use of the SIMBAD database, 
operated at CDS, Strasbourg, France.



\begin{thebibliography}

\bibitem{ack02} Acker, A., Gesicki, K., Grosdidier, Y., \& Durand, S. 2002, A\&A, 384, 620

\bibitem{ak90} Aaquist,  O. B., \& Kwok, S. 1990, A\&AS, 84, 229

\bibitem{bb93} Brand, J. \& Blitz, L. 1993, A\&A, 275, 67

\bibitem{cks92} Cahn, J. H., Kaler, J. B., \& Stanghellini, L.  1992, A\&AS, 94, 399

\bibitem{chs98} Christianto, H., \& Seaquist, E. R. 1998, AJ, 115, 2466 

\bibitem{grm93} G\'omez, Y., Rodr\'\i guez, L. F., \& Moran, J. M. 1993, ApJ, 416, 620.

\bibitem{ht96}  Hajian, A. R., \& Terzian, Y. 1996, PASP, 108, 419

\bibitem{htb93} Hajian, A. R., Terzian, Y., \& Bignell, C. 1993, AJ, 106, 1965

\bibitem{htb95} Hajian, A. R., Terzian, Y., \& Bignell, C. 1995, AJ, 109, 2600

\bibitem{km96}  Kawamura, J., \& Masson, C. 1996, ApJ, 461, 282 

\bibitem{lk05} Lee, T.-H., \& Kwok, S. 2005, ApJ, 632, 340 

\bibitem{lh98} Leuenhagen, U., \& Hamann, W.-R. 1998, A\&A, 330, 265

\bibitem{md92}  Maciel, W. J. \&  Dutra, C. M. 1992, A\&A, 262, 271

\bibitem{ma86}  Masson, C. R. 1986, ApJ, 302, L27

\bibitem{ma89a} Masson, C. R. 1989a, ApJ, 336, 294

\bibitem{ma89b} Masson, C. R. 1989b, ApJ, 346, 243 

\bibitem{me04} Mellema, G. 2004, A\&A, 416, 623

\bibitem{pa02}  Palen, S., Balick, B., Hajian, A. R., Terzian, Y., Bond, H. E., \& Panagia, N. 2002, AJ, 123, 2666

\bibitem{pms03} Pe\~na, M., Medina, S., \& Stasi\'nska, G. 2003, RMxAC, 18, 84 

\bibitem{re99} Reed, D. S., Balick, B., Hajian, A. R., Klayton, T. L., Giovanardi, S., Casertano, S., Panagia, N., \& Terzian, Y. 1999, AJ, 118, 2430 

\bibitem{stpp83} Schneider, S. E., Terzian, Y., Purgathofer, A., \& Perinotto, M. 1983, 
ApJS, 52, 399 

\bibitem{sch05} Sch\"onberner, D., Jacob, R., \& Steffen, M. 2005, A\&A, 441, 573


\end{thebibliography}
\end{document}